\documentstyle[12pt]{article}
\def\k{{\rm {\bf k}}}
\def\p{{\rm {\bf p}}}
\def\m{{m_{mag}^2}}

\begin{document}

\hfill BI-TP 96/09

\hfill February 1996

\vspace{1.5cm}

\begin{center}
{\bf MAGNETIC MASS IN HOT SCALAR ELECTRODYNAMICS }
\footnote{Research is partially supported by "Volkswagen-Stiftung"}

\vspace{1.5cm}
{\bf O.K.Kalashnikov}
\footnote{Permanent address: Department of
Theoretical Physics, P.N.Lebedev Physical Institute, Russian
Academy of Sciences, 117924 Moscow, Russia. E-mail address:
kalash@td.lpi.ac.ru}

Fakult\"at f\"ur Physik

Universit\"at Bielefeld

D-33501 Bielefeld, Germany

\vspace{2.5cm}
{\bf Abstract}
\end{center}

Using the Slavnov-Taylor identities we prove that the so-called
"magnetic mass" is exactly equal to zero within hot scalar
electrodynamics. The same result is valid for hot QED and seems 
for any abelian theory but this is not the case for hot QCD where 
one expects that $\m\ne 0$.

\newpage

At present for hot QCD and many other gauge theories it is very
essential to calculate the so-called "magnetic mass," which
is an infrared cutoff for gluomagnetic forces and in many cases it can
protect this theory from infrared divergencies. This question has a
very long history [1,2,3] but till now it is open to discussions. There
are only the estimates made perturbatively for this parameter [4,5,6]
although another possibility, which considers a nonanalytical behaviour
[7,8], is also not excluded. Nevertheless this parameter
(when $\m\ne 0$) is widely used today for many applications [9],
especially when the next-to-leading order term is calculated [10]
within hot QCD. Moreover it is often stated (starting from 
paper [11]) that for hot scalar electrodynamics and for any hot 
abelian theory this parameter is equal to zero although this fact has 
not been proven.

The goal of this paper is to calculate exactly the magnetic mass
for hot scalar electrodynamics using the Slavnov-Taylor identities.
Here we exploit the exact graph representation for the photon
self-energy tensor and demostrate that, indeed, this parameter is
equal to zero after the simple algebra being performed. Moreover we
also see arguments that this result is valid for hot QED and 
it is correct for any abelian theory. For hot QCD 
$\m\ne 0$ although the analogous calculations are also valid. On the 
formal level, the graphs with other numerical coefficients define the 
QCD self-energy tensor but, of course, the real reason is connected 
with the essential different nature of hot QCD infrared divergencies.

Scalar electrodynamics is determined through the Lagrangian
\setcounter{equation}{0}
\begin{eqnarray}
{\cal L}_A=-\frac{1}{4}F_{\mu\nu}^2-|(\partial_{\mu}-ieA_{\mu})\phi|^2
-\frac{\lambda}{4}(\phi^+\phi)^2
\end{eqnarray}
where $A_{\mu}$ is an abelian gauge field and $\phi^+(\phi)$ are the
complex scalar ones. Here $F_{\mu\nu}$ is the standard electromagnetic
field strength tensor and the last term in Eq.(1) is necessary to make
the model (1) renormalizable. The quantum Lagrangian for the theory
under consideration is built as usual and has the form
\begin{eqnarray}
{\cal L}={\cal L}_A+{\cal L}_{g.f.}\nonumber\\
{\cal L}_{g.f}=\frac{1}{2\alpha}(\partial_{\mu}A_{\mu})^2
+{\bar C}(\partial_{\mu}^2)C
\end{eqnarray}
where we add terms which fix the gauge and the appropriate ghost fields.

The set of equations for the temperature Green functions can be easily
obtained via the stationary-action principle [12] and has the standard
Schwinger-Dyson form
\begin{eqnarray}
D^{-1}(k_4,\k)=D_0^{-1}(k_4,\k)+\Pi(k_4,\k)\,,\,\,\,\,\,\
G(k_4,\k)=G_0^{-1}(k_4,\k)+\Sigma(k_4,\k)
\end{eqnarray}
where $\Pi$ and $\Sigma$ are the self-energy part of the photon Green
function and the Green function of scalar fields, respectively. The
explicit form of $\Pi$ can be represented by the four
nonperturbative graphs
\begin{eqnarray}
\vspace{4 cm}
\end{eqnarray}
where all lines and the bold points should be identified with the exact
Green and vertex functions. All the bare vertices are found to be
\begin{eqnarray}
\Gamma_{A\phi\phi^+}^0(k|p+k,p)_{\mu}=e(2p+k)_{\mu}\nonumber\\
\Gamma_{A^2\phi\phi^+}^0|_{\mu\nu}=-2e\delta_{\mu\nu}\,,\,\,\,\,\
\Gamma_{(\phi\phi^+)^2}^0=-\lambda
\end{eqnarray}
and they are independent from the gauge chosen. The last two functions
are independent from momenta as well.

For the Feynman gauge (where $\alpha=1$) the photon self-energy tensor
is transversal
\begin{eqnarray}
k_{\mu}\Pi_{\mu\nu}(k)=0
\end{eqnarray}
and can be represented with the aid of two scalar functions in the
form
\begin{eqnarray}
\Pi_{ij}(k_4,\k)&=&(\delta_{ij}-\frac{k_ik_j}{\k^2})A(k_4,\k)+
\frac{k_ik_j}{\k^2}\frac{k_4^2}{\k^2}\Pi_{44}(k_4,\k)\nonumber\\
\Pi_{i4}(k_4,\k)&=&\Pi_{4i}(k_4,\k)=-\frac{k_ik_4}{\k^2}\Pi_{44}(k_4,\k)
\,,\,\,\,\ i,j=1,2,3
\end{eqnarray}
The magnetic mass is determined to be
\begin{eqnarray}
\m=A(k_4=0,\k\rightarrow 0)
\end{eqnarray}
where the limit is defined in the infrared manner. However, since
$\Pi_{44}(k_4=0,\k\rightarrow 0)\ne 0$ for this theory, it is more
convenient to use for calculating $\m$ the relation
\begin{eqnarray}
\m=\frac{1}{2}\sum_i\Pi_{ii}(k_4=0,\k\rightarrow 0)
\end{eqnarray}
which directly follows from Eq.(7) when the Feynman gauge is used.

Our tool for transforming Eq.(4) is the exact Slavnov-Taylor identities
\begin{eqnarray}
\Gamma_{A\phi\phi^+}(0|p,p)_i&=&e\frac{\partial G^{-1}(p)}{\partial
p_i}\nonumber \\
\Gamma_{A^2\phi\phi^+}(0,k|p+k,p)_{ij}&=&-e
\frac{\partial \Gamma_{A\phi\phi^+}(k|p+k,p)_{i}}{\partial p_j}
\end{eqnarray}
which can be found by using the known prescription [12]. They are 
valid if one momentum is equal to zero in the infrared manner and for
indices $i,j\ne 4$.

One-loop nonperturbative graphs and two-loop ones in Eq.(4) are
canceled independently. The two first (one-loop) graphs are easily put
in the form
\begin{eqnarray}
\Pi_{ii}^{(1)}(0)=\frac{6e^2}{\beta}\sum_{p_4}\int\frac{d^3p}
{(2\pi)^3}G(p)-\frac{e^2}{\beta}\sum_{p_4}\int\frac{d^3p}{(2\pi)^3}
(2p_i)G(p)\frac{\partial G^{-1}(p)}{\partial p_i}G(p)
\end{eqnarray}
where we have used the first formula from Eq.(10) to eliminate the
$\Gamma_{A\phi\phi^+}$-function. Then using that $GG^{-1}=1$  
we arrive at the expression
\begin{eqnarray}
\Pi_{ii}^{(1)}(0)=\frac{6e^2}{\beta}\sum_{p_4}\int\frac{d^3p}{(2\pi)^3}
G(p)+\frac{e^2}{\beta}\sum_{p_4}\int\frac{d^3p}{(2\pi)^3}
(2p_i)\frac{\partial G(p)}{\partial p_i}
\end{eqnarray}
which is exactly equal to zero when one calculates the last integral by 
parts (the surface term being zero is ommited [13]). On this level we 
have that the nonperturbative one-loop $\Pi_{ii}^{(1)}(0)=0$. The same 
situation with the one-loop nonperturbative graphs takes place also in 
hot QCD which is possible to see, for example, in axial 
temporal gauge [13]. For hot QED the analogous calculations 
prove that $\m=0$ at once since the exact graph representation 
for the photon self-energy part in QED does not contain the 
nonperturbative two-loop graphs [12].

But there is a problem when the nonperturbative two-loop graphs 
are considered. For the model (1), however, we demostrate that the
two last nonperturbative graphs in Eq.(4) seem to be equal to zero as 
well. Here we take the third graph (below called $G_3$) from (4) which 
(after the first formula from Eq.(10) being used) has the form
\begin{eqnarray}
(G_3)=\frac{2e^3}{\beta^2}\sum_{k_4,p_4}
\int\frac{d^3k}{(2\pi)^3}\frac{d^3p}{(2\pi)^3}D_{ij}(k)G(p+k)
\Gamma_j(k|p+k,p)\frac{\partial G(p)}{\partial p_i}
\end{eqnarray}
and we perform the intergation by parts within Eq.(13). If the 
integral (below K-term)
\begin{eqnarray}
K=\frac{2e^3}{\beta^2}\sum_{k_4,p_4}
\int\frac{d^3k}{(2\pi)^3}\frac{d^3p}{(2\pi)^3}D_{ij}(k)
\left[\frac{\partial}{\partial p_i}G(p+k)
\Gamma_j(k|p+k,p)G(p)\right] 
\end{eqnarray}
is equal to zero, the expression Eq.(13) becomes
\begin{eqnarray}
(G_3)=-\frac{2e^3}{\beta^2}\sum_{k_4,p_4}
\int\frac{d^3k}{(2\pi)^3}\frac{d^3p}{(2\pi)^3}D_{ij}(k)
\left[\frac{\partial}{\partial p_i}G(p+k)
\Gamma_j(k|p+k,p)\right] G(p)
\end{eqnarray}
and this representation for $G_3$ is enough to prove within the 
model(1) that $\m=0$ exactly. Now one should explicitly perform 
a differentiation within Eq.(15) and find the simple identity
for the exact graphs within Eq.(4)
\begin{eqnarray}
\hspace{4 cm}
\end{eqnarray}
which shows that the magnetic mass for this model
is indeed equal to zero
\begin{eqnarray}
\m=0
\end{eqnarray}
However we should prove else that from Eq.(14) $K=0$.
In the lowest perturbative order (here this means the $e^4$-term) one 
can demonstrate that $K^{(0)}=0$, transforming Eq.(14) to the 
form
\begin{eqnarray}
K^{(0)}=\frac{2e^4}{\beta^2}\sum_{k_4,p_4}
\int\frac{d^3k}{(2\pi)^3}\frac{d^3p}{(2\pi)^3}\frac{1}{k^2}
\left[\frac{6}{(p+k)^2p^2}-\frac{8\p^2+4\p\k}{(p+k)^2p^4}\right] 
\end{eqnarray}
and then calculates it in the usual manner. For example, using the 
infrared manner of calculation, one finds at once that
\begin{eqnarray}
K^{(0)}=\frac{2e^4}{\beta^2}
\int\frac{d^3k}{(2\pi)^3}\frac{d^3p}{(2\pi)^3}\frac{1}{\k^2}\left[
-\frac{2}{(\p+\k)^2\p^2}-\frac{4\p\k}{(\p+\k)^2\p^4}\right]=0 
\end{eqnarray}
So there is not any problem with the leading $g^4$-term calculated 
for $\m$ and it being zero strongly
indicates that $\m=0$ within all perturbative orders.   
  
For hot QCD $\m\ne 0$ already within the $g^4$-order [4] although 
the analogous calculations are also possible, for example, in the 
axial temporal gauge. On the formal level the graphs with
other numerical coefficients define the QCD self-energy tensor but, 
of course, the real reason is connected with the essential different 
nature of the QCD infrared divergencies.

\newpage

\begin{center}
{\bf Acknowledgements}
\end {center}

I would like to thank Rudolf Baier for useful discussions and
all the colleagues from the Department of Theoretical Physics of the
Bielefeld University for the kind hospitality.

\begin{center}
{\bf References}
\end{center}

\renewcommand{\labelenumi}{\arabic{enumi}.)}
\begin{enumerate}

\item{ A.D.Linde, Phys. Lett. {\bf B 96} (1980) 289.}

\item{ D.J.Gross, R.D.Pisarski and L.G.Yaffe, Rev.Mod.Phys.
{\bf 53} (1981) 43.}

\item{ O.K.Kalashnikov, JETP Lett. {\bf 33} (1981) 165.}

\item{ O.K.Kalashnikov and E.Kh.Veliev, Sov. Phys. Lebedev Inst.
Rep. (1986) No.3, 39; for review see: O.K.Kalashnikov, 
Phys. Lett. {\bf B 279} (1992) 367.}

\item{ J.Blaizot, E.Iancu and R.R.Parwani, Phys. Rev. {\bf D 52}
(1995) 2543.}

\item{ R.Jackiw, So-Young Pi, Phys. Lett. {\bf B 368} (1996) 131. }

\item{ R.Jackiw and S.Templeton, Phys. Rev. {\bf D 23} (1981) 2291 .}

\item{ O.K.Kalashnikov, JETP Lett. {\bf 54} (1991) 181.}

\item{ R.D.Pisarski, Phys. Rev. {\bf D 47} (1993) 5589;
 F.Flechsig, A.K.Rebhan and H.Schulz, Phys. Rev. {\bf D 52}
 (1995) 2994.}

\item{ A.K.Rebhan, Phys. Rev. {\bf D 48} (1993) R3967;
E.Braaten and A.Nieto, Phys. Rev. Lett. {\bf 73} (1994) 2402.}

\item{ O.K.Kalashnikov and V.V.Klimov, Phys. Lett. {\bf B 95} 
(1980) 423.}

\item{ E.S.Fradkin, Proc. Lebedev Inst. 29 (1965) 6.}

\item{ O.K.Kalashnikov, JETP Lett. {\bf 39} (1984) 405.}

\end{enumerate}

\end{document}